%% file: nime-template.tex
\documentclass{nime-alternate} % Uncomment when publishing final version

\usepackage{graphicx, caption, subcaption}
\usepackage{float}
\usepackage{url}
\usepackage{anonymize} %                   Uncomment this line to publish
\input{math_commands.tex}

\begin{document}
%
% --- Author Metadata here ---
%\conferenceinfo{NIME'17,}{May 15-19, 2017, Aalborg University Copenhagen, Denmark.}
%\conferenceinfo{NIME'18,}{June 3-6, 2018, Blacksburg, Virginia, USA.}
\conferenceinfo{NIME'19,}{June 3-6, 2019, Federal University of Rio Grande do Sul, ~~~~~~  Porto Alegre,  Brazil.}
\title{Inspecting and Interacting with Meaningful Music Representations using VAE}

%
% You need the command \numberofauthors to handle the 'placement
% and alignment' of the authors beneath the title.
%
% For aesthetic reasons, we recommend 'three authors at a time'
% i.e. three 'name/affiliation blocks' be placed beneath the title.
%
% NOTE: You are NOT restricted in how many 'rows' of
% "name/affiliations" may appear. We just ask that you restrict
% the number of 'columns' to three.
%
% Because of the available 'opening page real-estate'
\label{key}% we ask you to refrain from putting more than six authors
% (two rows with three columns) beneath the article title.
% More than six makes the first-page appear very cluttered indeed.
%
% Use the \alignauthor commands to handle the names
% and affiliations for an 'aesthetic maximum' of six authors.
% Add names, affiliations, addresses for
% the seventh etc. author(s) as the argument for the
% \additionalauthors command.
% These 'additional authors' will be output/set for you
% without further effort on your part as the last section in
% the body of your article BEFORE References or any Appendices.

\numberofauthors{4} %  in this sample file, there are a *total*
% of EIGHT authors. SIX appear on the 'first-page' (for formatting
% reasons) and the remaining two appear in the \additionalauthors section.
%
\author{
% You can go ahead and credit any number of authors here,
% e.g. one 'row of three' or two rows (consisting of one row of three
% and a second row of one, two or three).
%
% The command \alignauthor (no curly braces needed) should
% precede each author name, affiliation/snail-mail address and
% e-mail address. Additionally, tag each line of
% affiliation/address with \affaddr, and tag the
% e-mail address with \email.
%
% 1st. author
\alignauthor
\anonymize{Ruihan Yang}\\
      \affaddr{\anonymize{Music X Lab}}\\
      \affaddr{\anonymize{NYU Shanghai}}\\
      \email{\anonymize{ry649@nyu.edu}}
% 2nd. author
\alignauthor
\anonymize{Tianyao Chen}\\
      \affaddr{\anonymize{Music X Lab}}\\
      \affaddr{\anonymize{NYU Shanghai}}\\
      \email{\anonymize{tc2709@nyu.edu}}
% 3rd. author
\alignauthor \anonymize{Yiyi Zhang}\\
      \affaddr{\anonymize{Center for Data Science}}\\
      \affaddr{\anonymize{New York University}}\\
      \email{\anonymize{yz2092@nyu.edu}}
 % use '\and' if you need 'another row' of author names
% 4th. author
\alignauthor \anonymize{Gus Xia}\\
      \affaddr{\anonymize{Music X lab}}\\
      \affaddr{\anonymize{NYU Shanghai}}\\
      \email{\anonymize{gxia@nyu.edu}}
% 5th. author
% \alignauthor \anonymize{Sean Fogarty}\\
%       \affaddr{\anonymize{NASA Ames Research Center}}\\
%       \affaddr{\anonymize{Moffett Field}}\\
%       \affaddr{\anonymize{California 94035}}\\
%       \email{\anonymize{fogartys@amesres.org}}
% % 6th. author
% \alignauthor \anonymize{Anon Nymous}\\
%       \affaddr{\anonymize{Redacted }}\\
%       \affaddr{\anonymize{8600 Datapoint Drive}}\\
%       \affaddr{\anonymize{San Antonio, Texas 78229}}\\
%       \email{\anonymize{cpalmer@prl.com}}
}
% There's nothing stopping you putting the seventh, eighth, etc.
% author on the opening page (as the 'third row') but we ask,
% for aesthetic reasons that you place these 'additional authors'
% in the \additional authors block, viz.
% \additionalauthors{Additional authors: \anonymize{John Smith (The Th{\o}rv{\"a}ld Group,}
% email: {\texttt{\anonymize{jsmith@affiliation.org}}}) and \anonymize{Julius P.~Kumquat 
% (K. Consortium,} email: {\texttt{\anonymize{jpkumquat@consortium.net}}}).}
% \date{30 July 1999}
% Just remember to make sure that the TOTAL number of authors
% is the number that will appear on the first page PLUS the
% number that will appear in the \additionalauthors section.

% For your initial submission you MUST ANONYMIZE the authors.

\maketitle 
\begin{abstract}
Variational Autoencoders~\cite{kingma2013auto} (VAEs) have already achieved great results on image generation  and recently made promising progress on music generation. However, the generation process is still quite difficult to control in the sense that the learned latent representations lack meaningful music semantics. It would be much more useful if people can modify certain music features, such as rhythm and pitch contour, via latent representations to test different composition ideas. In this paper, we propose a new method to inspect the pitch and rhythm interpretations of the latent representations and we name it \textit{disentanglement by augmentation}. Based on the interpretable representations, an intuitive graphical user interface is designed for users to better direct the music creation process by manipulating the pitch contours and rhythmic complexity.
\end{abstract}

\keywords{Representation learning, Disentanglement, Music generation, Controlled generation}

\section{Introduction}
Representation learning has become an essential tool to gain an in-depth view of the data. Bengio~\cite{bengio2013representation} pointed out that good data representations ``make it easier to extract useful information when building classifiers or other predictor". We have also seen that representation learning dramatically boosted the effectiveness of generative models for visual arts and music style transfer~\cite{DBLP:journals/corr/abs-1803-06841, DBLP:journals/corr/abs-1809-07600}. In particular, the general encoder-decoder architecture of Variational Autoencoders (VAEs)~\cite{kingma2013auto} (and in the same sense, the generator-discriminator architecture of Generative Adversial Network~\cite{goodfellow2014generative}) provides a way to generate data by sampling from the distribution of the latent representation, rather than directly sampling from the data distribution or generating one token or pixel at one time. More recently, the pioneer work of MusicVAE~\cite{roberts2017hierarchical} incorporated sequence modeling with VAEs by building both the encoder and decoder with Long Short-Term Memory networks (LSTMs)~\cite{hochreiter1997long}. However, since the learned latent representations lack semantic interpretations, the generation process is still quite difficult to control. From a practical standpoint, it would be helpful if users could manipulate meaningful music features via latent representations during the music creation process in a similarly way of tuning the radio with turning knobs, in order to test different composition ideas efficiently.

We aim to gain a better semantic interpretation of the learned latent representations by disentanglement, i.e., to inspect which part (dimensions) of the representations connects with which features of music composition, while providing an intuitive interface to control the disentangled features. In this paper, we adopt the MusicVAE model~\cite{roberts2017hierarchical} and focus on disentangling and interacting with \textit{pitch contour} and \textit{rhythm} representations of symbolic melodies. We propose a new method named \textit{disentanglement by augmentation} and conduct the experiment in two ways: one way only transpose the pitch contour, and the other way only split the note duration, both gradually. We discover that encoded latent representations change approximately linearly along with the changing of pitch contour and rhythm. Moreover, only a small portion of dimensions changes significantly comparing to the entire large space. Among them, almost no dimension contributes to both pitch contour changes and rhythm changes. Therefore, pitch contour and rhythm are considered to be disentangled. By fixing one of these sets of dimensions and releasing the other one, we can make the original music transformed to new music in a desired way. Meanwhile, we provide an intuitive interface for users to create music via directly interacting with or interpolating the significant dimensions of pitch contour and rhythm representations. Available at: \url{https://github.com/cdyrhjohn/representation_demo}

In the rest of the paper, Section \ref{sec:background} gives a background knowledge about the VAEs and MusicVAE models, Section \ref{sec:method} discusses how to inspect the latent space, Section \ref{sec:controlled-interaction} introduces the interface we design for music composition, and the last section summaries our study and discusses possible future work.

\section{Background}
\label{sec:background}
Our study is built on VAEs~\cite{kingma2013auto}. In this section, we review how VAEs work and how to adjust the model architecture to deal with the representation learning of symbolic music sequences.

\subsection{Variational Autoencoder}
The VAE~\cite{kingma2013auto} is a classical generative model that can learn a low-dimensional latent vector $\vz$ from a high-dimensional data $\vx$. The latent vector $\vz$ can be used to generate new data or to reconstruct the original data. The prior distribution of $\vz$ is $p(\vz)$. Therefore, the generated data $\vx$ need to satisfy the distribution  $\vx\sim p(\vx|\vz)$. The encoder in the VAE uses a distribution $q(\vz|\vx)$ to approximate $p(\vz|\vx)$ and the decoder parameterizes the distribution $p(\vx|\vz)$.  The objective is to minimize the KL divergence between $q(\vz|\vx)$ and $p(\vz)$ by maximizing the evidence lower bound:
\begin{equation}
    \mathbb{E}[\log{p(\vx|\vz)}] - \text{KL}(q(\vz|\vx)\ ||\ p(\vz)) \leq \log{p(\vx)}
\end{equation}
Generally speaking, both encoder and decoder are neural networks, and $p(\vz)$ is parameterized as a diagonal covariance Gaussian, i.e., $\vz\sim\mathcal{N}(\mathbf{\mu}, \mathbf{\sigma})$.

In our model, after the training period, we take the mean of distribution $p(\vz)$, making $\vz = \mathbf{\mu}$, as the latent representation of the input data.

\subsection{Music Variational Autoencoder}
\begin{figure}[H]
	\centering
	\includegraphics[width=0.85\linewidth]{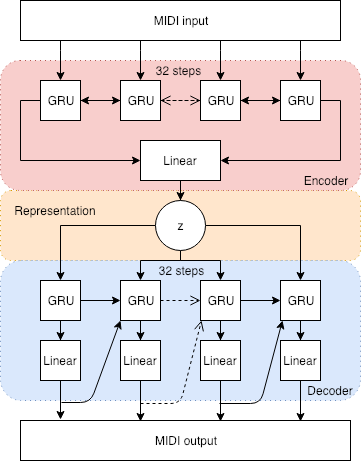}
	\caption{An illustration of the VAE model to learn music representation}
    \label{fig:vae}
\end{figure}

MusicVAE~\cite{roberts2017hierarchical} is an application of variational autoencoder on music. The format of the input music required by this model is in the form of a 2-bar MIDI sequence with a 4/4 meter. Each 2-bar music clip is represented by a $32\times130$ matrix, with 32 time steps in the unit of sixteenth note, and 130 states on each time step, including 128 onset states for each MIDI pitch, a holding state, and a rest state. To deal with this time-series data, MusicVAE uses bidirectional Long Short-Term Memory networks (LSTMs)~\cite{hochreiter1997long} recurrent neural networks for both encoder and decoder. In this paper, we use Gated Recurrent Units (GRUs)~\cite{chung2014empirical} as a replacement of LSTMs for more efficient training processes. The model architecture is shown in Figure \ref{fig:vae}. The learned low-dimensional latent vector $\vz$ can be seen as a compact and continuous latent representation, based on which we can smoothly transform from one music clip to another one through interpolation or adding certain features via attribute vector arithmetic~\cite{hinton2015distilling}.

Note that the original MusicVAE model has a hierarchical structure to handle longer sequences, but results on music with two bars are not yet convincing. Therefore, we only adopt the model to learn short sequences. When dealing with longer sequences, we cut them into 2-bar clips, process them one by one, then concatenate them together.
\vspace{2pc}

\section{Methods}
\label{sec:method}
Our goal is to allow music composers to manipulate latent representations of a music melody easily with an effective and intuitive interface. Two improvements are needed based on MusicVAE~\cite{roberts2017hierarchical}. First, we need a more compact latent representation. To this end, we shrink the dimensionality of the latent space from 512 to 128. Second, we need to disentangle the latent representation in a meaningful way. Each disentangled part of the representation should coincide with some explicit music concept, such as pitch or rhythm, thus human composers can manipulate these music features by changing the values on the latent representations while knowing the consequences. As an analogy, the sound produced by a radio can be controlled by tuning labelled knobs on a radio, while each label, such as volume or frequency, assigned to the knobs tells users the consequences of tuning them. The meaningful disentangled representation makes the design of the interface more effective and efficient for human interaction and testing creative composition ideas.

% To better understand, we can compare the MusicVAE model as a radio, and its output as the wave we hear from the radio. After completing the disentanglement, we have equipped the model with 2 turning knobs: one controls the key of the music and the other controls the note density. As a result, users now can test their ideas by easily tuning those 2 features just as tuning the knobs on the radio to find their favorite broadcasting station.

To help the model extract representations more precisely, the KL divergence is re-weighted by a value $\beta$, which is set as 0.1 in our case to help the model focus more on reconstruction rather than innovation. We refer readers to~\cite{higgins2016beta} for more information on $\beta$-VAE.

\begin{figure}[H]
	\centering
	\includegraphics[width=\linewidth]{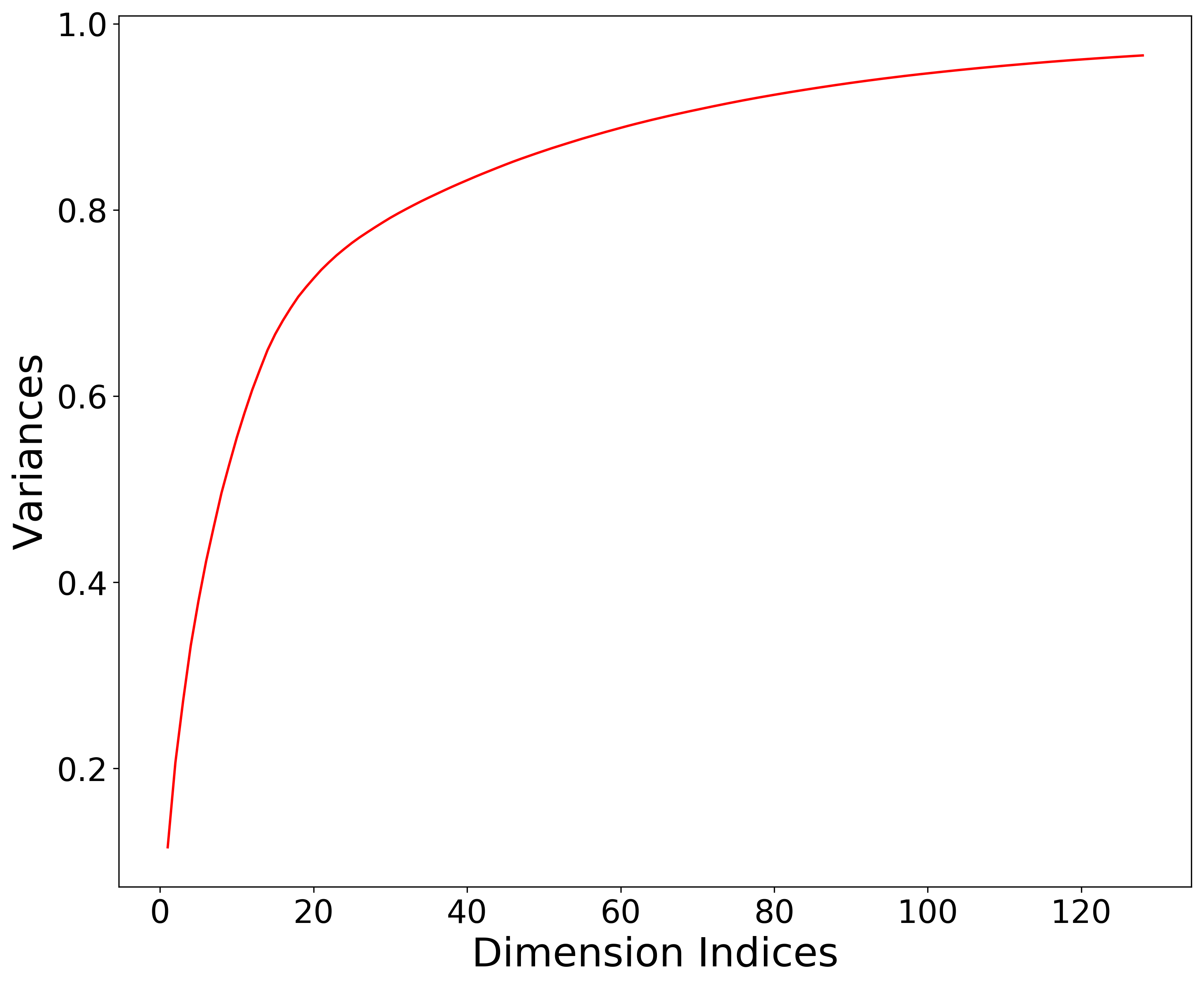}
	\caption{Sorted variance in the 128-dimension vector after PCA}
	\label{fig:pca}
\end{figure}

\subsection{A More Compact Representation}
We first examined the size of the latent space. The dimensionality of the latent space designed by the original MusicVAE~\cite{roberts2017hierarchical} is too large for an intuitive and controllable interface. MusicVAE uses a Euclidean space with 512 dimensions. We train a MusicVAE model on our dataset, of which the encoder maps each music clips to a 512-dimension vector in the latent space.

We performed \textit{principal components analysis (PCA)}~\cite{wold1987principal} on latent vectors of music clips across the dataset. As shown in figure \ref{fig:pca}, we observe that more than $99\%$ of the aggregated eigenvalues are covered by the first 128 dimensions. Meanwhile, the sorted variance in figure \ref{fig:pca} indicates that even less than 128 dimensions have major variances. To be cautious, we select 128 as our latent space size to ensure that the model could learn a more complete representation. In the following sections, we use a MusicVAE with only 128 latent space dimensions pretrained on the dataset.

% \ruihan{The latent representation in vanilla MusicVAE is too large for manual inspection, so we believe its necessary to reduce the corresponding dimensionality. We leverage \textit{principle components analysis (PCA)} ~\cite{wold1987principal} to validate and achieve dimensionality reduction, which brings a more human-readable representations in our inspection and interaction.}
% \gus{font of the picture should be much bigger. Derek, please step in.}

% Therefore, an essential point in our model setup is that only 128-dim vector is used as our latent representation. Before we use this parameter setup, we leverage PCA on a batch of 512-dim representations learnt from the 100k 2-bar music clips with a pre-trained model and the result shows that 128 over 512 dimensions can reflect more than 99\% information, which indicates the potential of lowering the dimensionality. We also plot and sort the variance after the dimensionality reduction is applied provide a stronger evidence demonstrates that the 128 dimensions can be well enough to cover the information in 2-bar music clips, which also indicates that major variances are covered within 100 dimensions (see Figure \ref{fig:pca}). Therefore, a model is retrained with 128-dim latent representation to help explain our further analysis.

\subsection{Disentanglement by Augmentation}
\label{sec:D-by-A}

We propose a new method named \textit{disentanglement by augmentation}, which can be considered a special case of \textit{analysis by synthesis}~\cite{yuille2006vision}.
Start from a well-trained MusicVAE, of which the training process follows the original MusicVAE paper~\cite{roberts2017hierarchical}, which consists of an $\text{ENCODER}$ function and a $\text{DECODER}$ function. The $\text{ENCODER}$ maps a certain observed music clip $\tM_i$ to a latent vector $\vz_i$, while the $\text{DECODER}$ maps the latent vector $\vz_i$ back into the original music clip $\tM_i$.

Now, consider a data augmentation function $F$, which can directly transform the input music clips. Theoretically, for any $F$, there would be a corresponding function $f$ in the latent space, which moves the latent vector of the original music clip to the latent vector of the transformed music clip. Formally,
\begin{equation}
    f(\text{ENCODER}(\tM_i)) = \text{ENCODER}(F(\tM_i))
\end{equation}

For a latent vector $\vz_i$ of music clip $\tM_i$, transformation $f$ can be in the form of
\begin{equation}
    f(\vz_i) = \vz_i + \Delta_f \vz_i.
\end{equation}
Since $p(\vz)$ satisfies a multivariate independent normal distribution, an ideal assumption about $f$ is that, given a fixed transformation $F$ and its corresponding latent space transformation $f$, for all music clips $\tM_i$, the differences between latent vectors before and after the transformation, $\Delta_f \vz_i$, are non-zero in certain components, while the rest are kept zero.

The assumption is obviously too strong. Instead, we have another practical hypothesis that, latent vectors have significant changes only on a few dimensions. Given an orthonormal basis of the latent space, $\{\ve_1,\ve_2,\ldots,\ve_n\}$, we assume that
\begin{equation}
    \left| \Delta_f \vz_i \cdot \ve_{d_j} \right| \geq t
\end{equation}
on dimensions $\{d_1,d_2,\ldots,d_k\}$, where $t$ is a threshold determined manually.

We further assume that on those dimensions, latent vectors have significant changes on the same direction, either positive correlated or negative correlated. Therefore, we can use the average of the differences to select the dimensions of significant change,
\begin{equation}
    \left| \overline{\Delta_f \vz} \cdot \ve_{d_j} \right| \geq t \quad \overline{\Delta_f \vz} = \sum_{i=1}^N \Delta_f \vz_i
\end{equation}

For simplicity, we choose the standard basis as our orthonormal basis of the latent space.

\subsection{Inspecting Pitch Representations}
\label{sec:pitch-repr}

To inspect which dimensions of the latent space contribute the most to pitch information, we defined a set of pitch augmentation function $F^{\text{pitch}}_{p}$, which transposes the pitches of all notes of a music clip up by $p$ semitones . (For instance, $F^{\text{pitch}}_{3}$ means to transpose the pitches up by a minor third.) For each $p$, we calculated a set of $\Delta \vz^{p}_i$ and its average $\overline{\Delta_f \vz^p}$. In practice, a batch size of 10k is used, and $p$ ranges from 1 (a half step) to 12 (an octave). 

\begin{figure}[htbp]
    \centering
	\includegraphics[width=\linewidth]{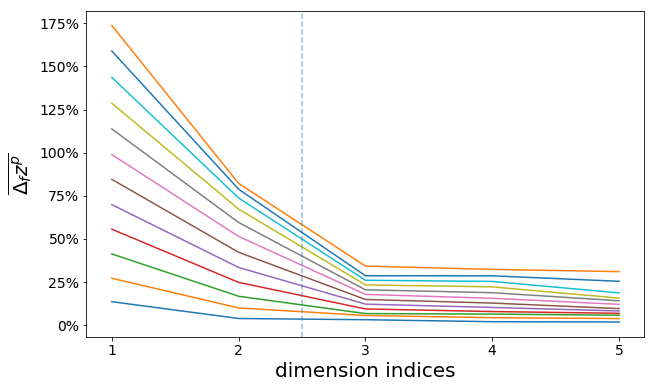}
	\caption{An illustration of only two latent dimensions contribute the most to pitch variation.}
	\label{fig:pitch-analysis}
\end{figure}

Figure \ref{fig:pitch-analysis} shows the top five latent dimensions that change most for different $p$, where the twelve lines corresponds to the twelve values of $p$. We see that, for each $p$, only 2 dimensions of the latent representations change significantly, while other dimensions almost keep the same. In addition, the sets of significant dimensions for different $p$ overlap a lot, and the top 2 dimensions always remain the same for different $p$. This discovery supports our assumption in section \ref{sec:D-by-A}. We conclude that only a small portion (only two dimensions) of the latent representations contributes to pitch variation, and we refer to these two dimensions as \textit{pitch representations}.

% Figure \ref{fig:pitch} illustrates the exact average value changes on the latent representations after we move up the pitch for 1 semitone, where the 2 dimensions with highest absolute values correspond the top 2 dimensions in Figure \ref{fig:pitch-analysis}.

% \begin{figure}[H]
%     \centering
%     \includegraphics[width=\linewidth]{imag/dim_change_p}
%     \caption{Value change on each dimension (pitch)}
%     \label{fig:pitch}
% \end{figure}

\subsection{Inspecting Rhythm Representations}
\label{sec:rhythm-repr}

To inspect which dimensions of the latent space contribute the most to rhythm information, we define a set of rhythm augmentation function $F^{\text{rhythm}}_{n}$, which splits each note into $n$ equal-length ones with the same pitch. For instance, $F^{\text{rhythm}}_{2}$ splits a quarter note into two consecutive eighth notes. This augmentation cuts the melody but the rough pitch trend still remains. For each $n$, we get a set of $\Delta z^n_i$ and its average $\overline{\Delta_fz^n}$. In practice, we build a dataset which has 16483 $2$-bar music clips, each is composed of two whole notes. For $n$, we take $n=2$, $n=4$, $\ldots$, $n=16$.

\begin{figure}[htbp]
    \centering
    \includegraphics[width=\linewidth]{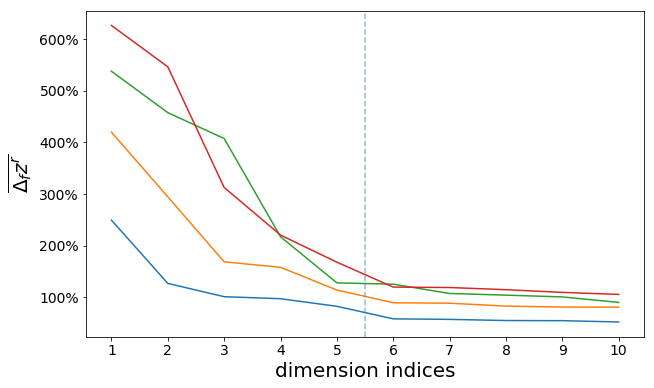}
    \caption{An illustration of only five latent dimensions contribute the most to rhythm variation}
    \label{fig:rhythm-analysis}
\end{figure}

Figure \ref{fig:rhythm-analysis} shows the $10$ most significant dimensions in the latent space. Curves with different color corresponds to different values of $n$. We observe that the influence of different $n$ may vary from dimensions to dimensions but the most significantly changed ones concentrate in only $5$ dimensions, while other dimensions almost keep intact. This discovery proves our assumption in section \ref{sec:D-by-A}. We conclude that only a small portion of the latent representation contributes to rhythm variation, and we refer to these dimensions as \textit{rhythm representations.}

% Figure \ref{fig:rhythm} illustrates the average value change in $\Delta \vz$ after we split the whole-note clips to half-note clips with the similar scenario as moving the pitch.% % \begin{figure}]
% % 	\centing
% % 	\includegraphics[width=\linewidth]{imag/dim_chge_r}
% % 	\caption{Value change on each dimension hythm)}
% % 	\label{g:rhythm}
% % nd{figure}
% \gus{checkpoint: looks good so far, only needs minor modification}

\subsection{Pitch-Rhythm Disentanglement}

In sum, the top 2 dimensions selected in Section \ref{sec:pitch-repr} are referred to as the latent pitch representation, and the top 5 dimensions selected in Section \ref{sec:rhythm-repr} are referred to as the latent rhythm representation. We observe that there is no overlap between these two groups of representations.

In our following experiments of interactive composition, pitch and rhythm are modulated only via their corresponding latent representations, while keeping other dimensions fixed.

\section{Controlled Interaction}
\label{sec:controlled-interaction}
Latent space representations encode full information of original music clips, so that modulating latent vectors will result in changes of music clips. Based on the pitch-rhythm disentanglement achieved in section \ref{sec:method}, we can interact with the music representation in a human-interpretable and non-trivial ways.

\begin{figure}[htbp]
    \centering
    \includegraphics[width=\linewidth]{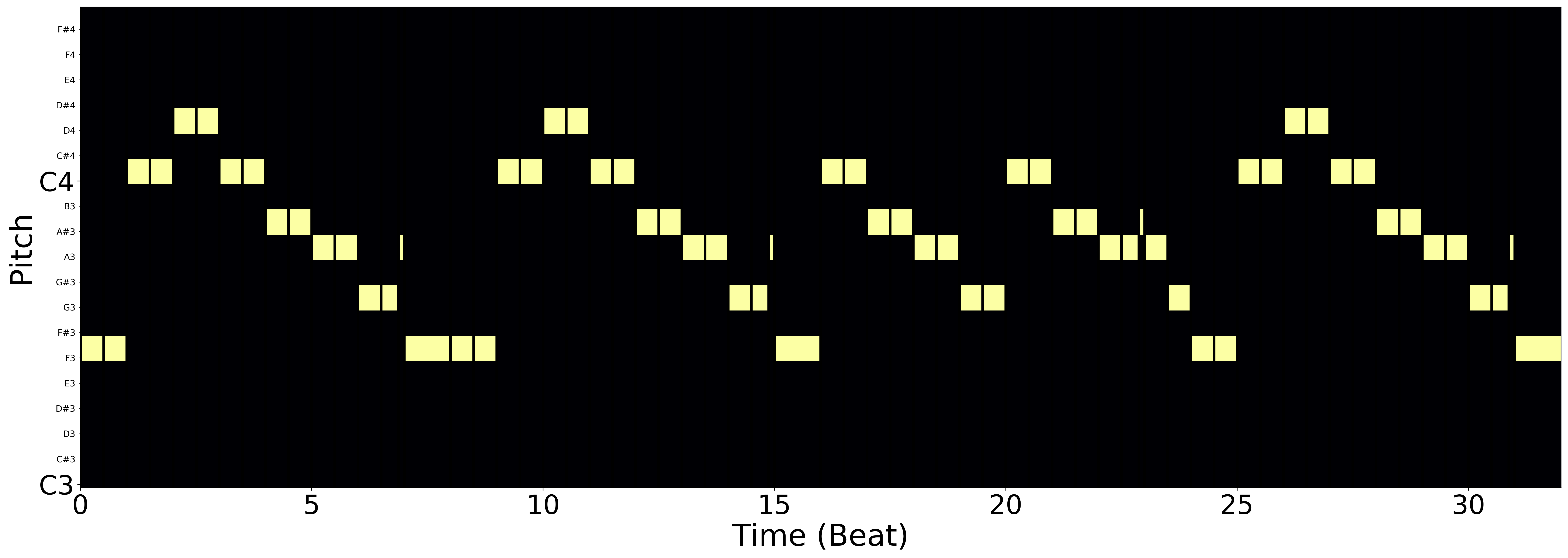}
    \caption{The piano representation of the original sample.}
    \label{fig:xxx}
\end{figure}

In this section, we use the theme of the ``Twelve Variations on `Ah vous dirai-je, Maman''' (Twinkle, Twinkle, Little Star)
by Mozart to illustrate the process of the human-computer interactive composition by controlling the pitch and rhythm representations. Figure \ref{fig:xxx} shows the piano roll of this music theme.
% \gus{minor revision, plz check.}
% \gus{make the fig shorter, but make the font much much bigger. We should be able to tell the pitch range from the y-label, rather than from the black area}
% In this section, The interaction methods we used for music creation with latent representation will be introduced.

\subsection{Pitch Interaction}
\label{sec:pi}
By performing transformation on pitch representation introduced in section \ref{sec:pitch-repr}, we will get the most significant dimensions of the latent space for pitch variation. To create new music via precisely interacting with the pitch representation of the original music, the interface performs the following operations:
\begin{enumerate}
    \item cut the original music into consecutive 2-bar clips, 
    \item encode the music clips into latent representation vectors,
    \item for each latent vector, only modulate the pitch representations (the most significant dimensions for pitch variation),
    \item decode the modified representations back to new music clips, and
    \item concatenate the new clips to form a new piece of music.
\end{enumerate}
% \gus{Any method to control the space between the items?}
Figure 6(a) displays the new music theme created via increasing the values of the 2-dimensional pitch representation of the original theme, while figure 6(b) is another created theme by applying the same amount of increment on 2 random dimensions of the latent representation that neither belong to pitch nor rhythm representation.

% \gus{alright, I am imagining here, but you help me realize it}

We can see that the new piece 6(a) significantly increases the overall pitch registration based on the original sample without much modification on the rhythm. In comparison, piece 6(b) hardly changes anything of the original theme. This difference indicates that the our pitch disentanglement is successful. Moreover, piece 6(a) is not merely transposing the original pitches but also creating a new melody with higher pitches and a slightly different pitch contour. An audio version of piece 6(a) can be found at \url{soundcloud.com/user-705441005/increase-pitch}. This discovery indicates that the interface successfully helps us generate new melodic ideas by controlling the pitch representation.

\begin{figure}[htbp]
    \centering
    \begin{subfigure}{\linewidth}
    \includegraphics[width=\linewidth]{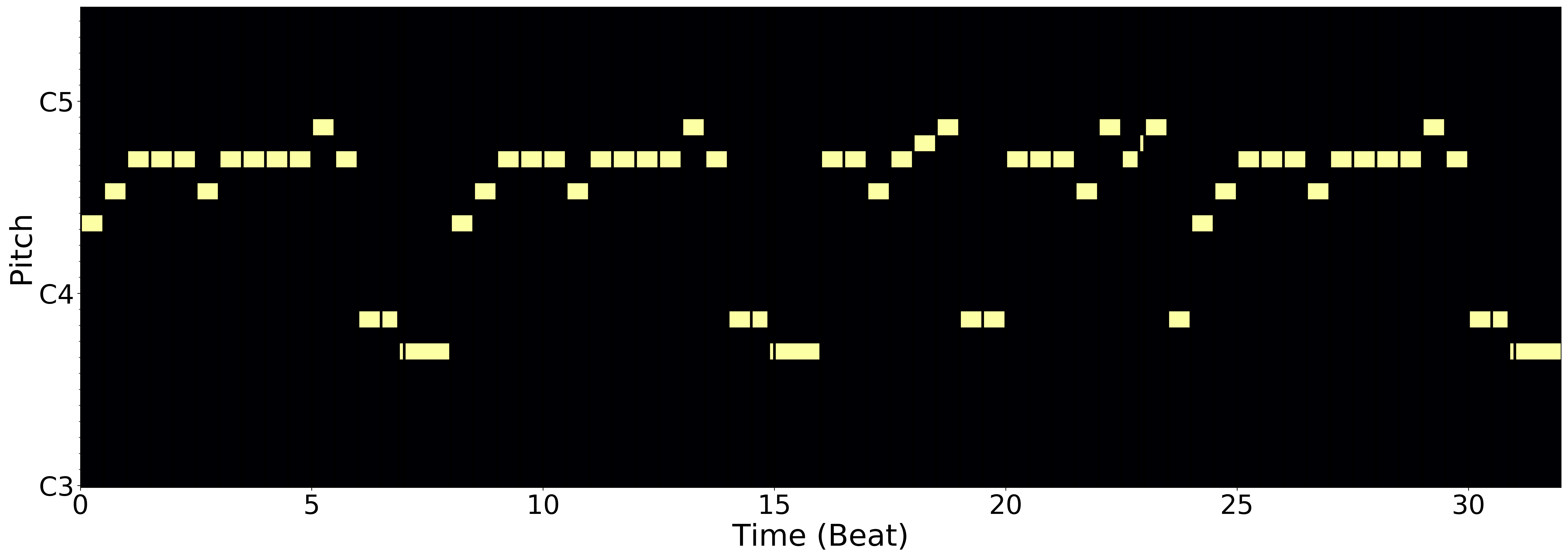}
    \caption{An illustration of interactive composition via controlling pitch representation.}
    \label{fig:xxx-p}
    \end{subfigure}
    \begin{subfigure}{\linewidth}
    \includegraphics[width=\linewidth]{imag/xxx}
    \caption{An illustration of interactive composition via controlling 2 random unrelated representation dimensions.}
    \label{fig:xxx-pr}
    \end{subfigure}
    \caption{Interactive composition on pitch.}
\end{figure}

% \gus{major revision, plz check!}

% \ruihan{plz move the figures below to the correct position}

% \yiyi{need a figure}
% \gus{we need an example of "xxx"}

% Based on the pitch representations, i.e., the significant dimensions of the latent representation contributing to the pitch variation, we can precisely interact with the pitch content of different music for our own composition.  \gus{we need an example of "xxx"}

% Note that the new music is not merely transposing the original music, but also creating new melody... 

% In order to pick out a set of the most effective dimensions, we also did a data visualization to rank all the changes that took place on every dimension. Within each group, gaps will show after listing nearly 2 dimensions. The absolute numbers on other dimensions seem more minor. Therefore, we have confidence to say that those 2 dimensions are actually taking charge of one of important features of the music clip, pitch.

\subsection{Rhythm Interaction}
\label{sec:ri}

To create new music via precisely interacting on the rhythm representation of the original music, the interface performs almost the same five steps as shown in section \ref{sec:pi}. The only difference lies in the third step, where the focus now shifts to the rhythm representation.

% By performing transformation on rhythm representation introduced in section \ref{sec:rhythm-repr}, we will get the most significant 5 dimensions on the latent space on rhythm division. By implementing the following operation, we will get a new music clip, of which notes would be splitted,

% \begin{enumerate}
%     \item encode the music clip into a vector on the latent space,
%     \item increase components of the disentangled latent representation on the most significant 5 dimensions on rhythm, and
%     \item decode the modified representation by the decoder.
% \end{enumerate}

% The greater the number of divisions of notes desired to be splitted, the greater the values on the second step needed to be increased.

% This operation is also a mapping from music clips to music clips. The output music shows higher note density than the input one. Although the values increased on the disentangled latent representation are obtained by inspection on datasets that equally and uniformly divide notes, this operation performs an irregular change on rhythm.

\begin{figure}[htbp]
    \centering
    \begin{subfigure}{\linewidth}
    \includegraphics[width=\linewidth]{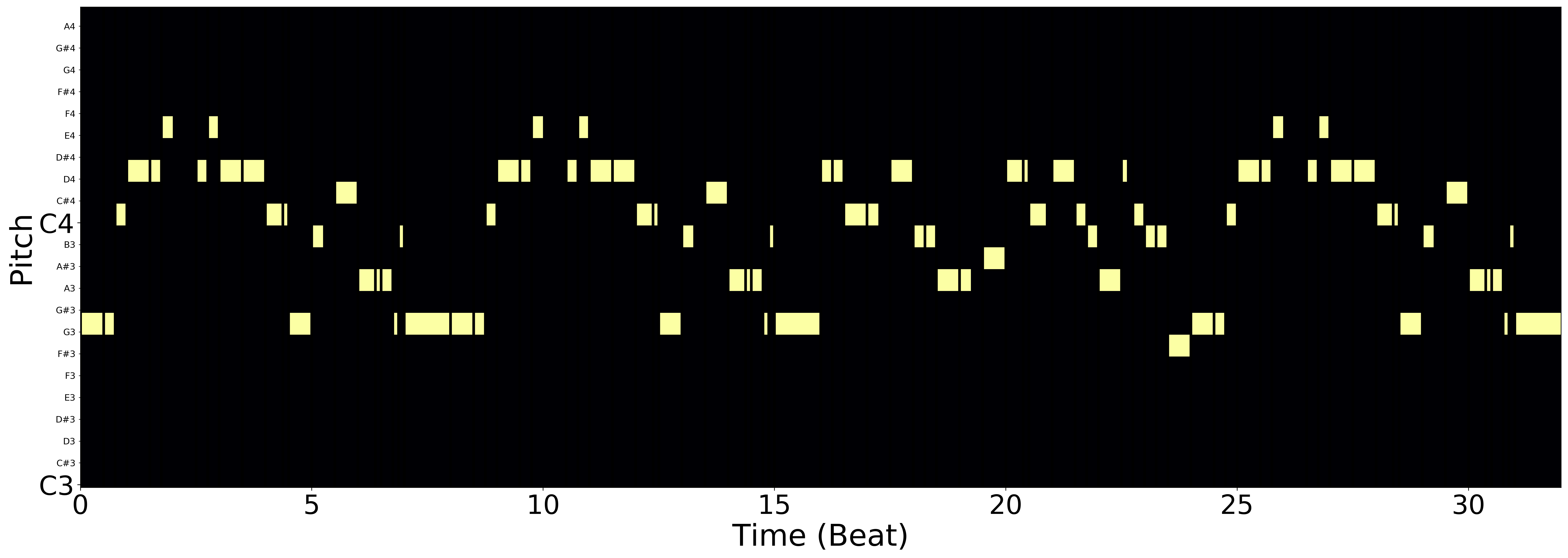}
    \caption{An illustration of interactive composition via controlling rhythm representation.}
    \label{fig:xxx-r}
    \end{subfigure}
    \begin{subfigure}{\linewidth}
    \includegraphics[width=\linewidth]{imag/xxx}
    \caption{An illustration of interactive composition via controlling 5 random unrelated representation dimensions}
    \label{fig:xxx-rr}
    \end{subfigure}
    \caption{Interactive composition on rhythm.}
\end{figure}

Figure 7(a) displays the new music theme created via increasing the values of the 5-dimensional rhythm representation of the original theme, while figure 7(b) is another created theme by applying the same amount of increment on 5 random dimensions of the latent representation that neither belong to pitch nor rhythm representation.

% \gus{alright, I am imagining here, but you help me realize it}

We see that the rhythm of new piece 7(a) is significantly changed. Another observation is that the pitches are not exactly the same as before. Instead, the pitches change smoothly and follow the original melody trend. This result is reasonable because the disentanglement is meant to happen at the representation level, not the observation level. The smooth transfer also indicates that our model has comprehended the intrinsic nonlinear relationship between pitch and rhythm and thus merges them in an organic way. In comparison, piece 7(b) hardly changes anything of the original theme. This difference indicates that the our rhythm disentanglement is successful. Moreover, the algorithm does not merely cut the notes to produce the new piece 7(a). Instead, it creates a new melody that retains the original melody contour. The music actually sounds like an electronic game version of the original theme, whose audio can be found at \url{soundcloud.com/user-705441005/increase-note-density}. This discovery indicates that the interface successfully helps us generate new melodic ideas by controlling the rhythm representation.
% \gus{major revision, plz check!}
% Figure xxx displays the comparisons when adding values to significant or non-significant dimensions. The top piano roll represents the original music, the middle piano roll is the output we get by adding the most significant xxx dimensions and the bottom piano roll shows the music clip we get if we add values on 100 unimportant dimensions.

% Likewise, we can precisely interact with the rhythm content of different music for our own composition.  \gus{we need one/two example of "xxx", the eletronic one}

\subsection{Extra Interaction}

\begin{figure}[htbp]
    \centering
    \includegraphics[width=\linewidth]{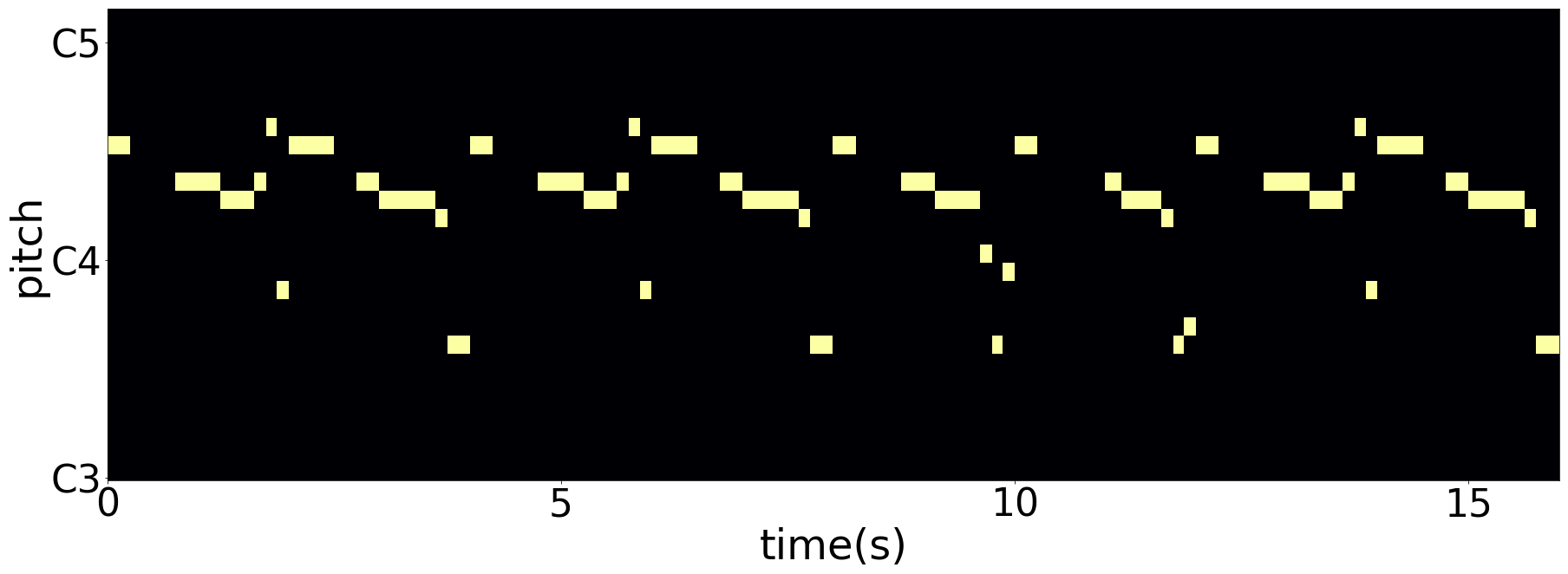}
    \caption{An illustration of interactive composition via controlling non-significant representations}
    \label{fig:xxx-rest}
\end{figure}

Besides interacting with pitch and rhythm representations, which leads to interpretable results, we provide an option to modulate the latent dimensions that do not have significant impacts either on pitch or on rhythm. Note that non-significance is not equivalent to uselessness. They indeed contribute to data reconstruction but just in a manner which is hard to be interpreted. Figure \ref{fig:xxx-rest} shows the result via increasing the value of the non-significant latent dimensions. We see that both melody contour and rhythm patterns are modified, but the result is far less musical compared to ones shown in figure \ref{fig:xxx-r} and figure \ref{fig:xxx-p}. The audio file can be found at \url{soundcloud.com/user-705441005/rest-dim-modulate}.
% \gus{major revision, plz check}

\subsection{Two-way Style-transfer Interpolation}
\begin{figure}[htbp]
	\centering
	\includegraphics[width=\linewidth]{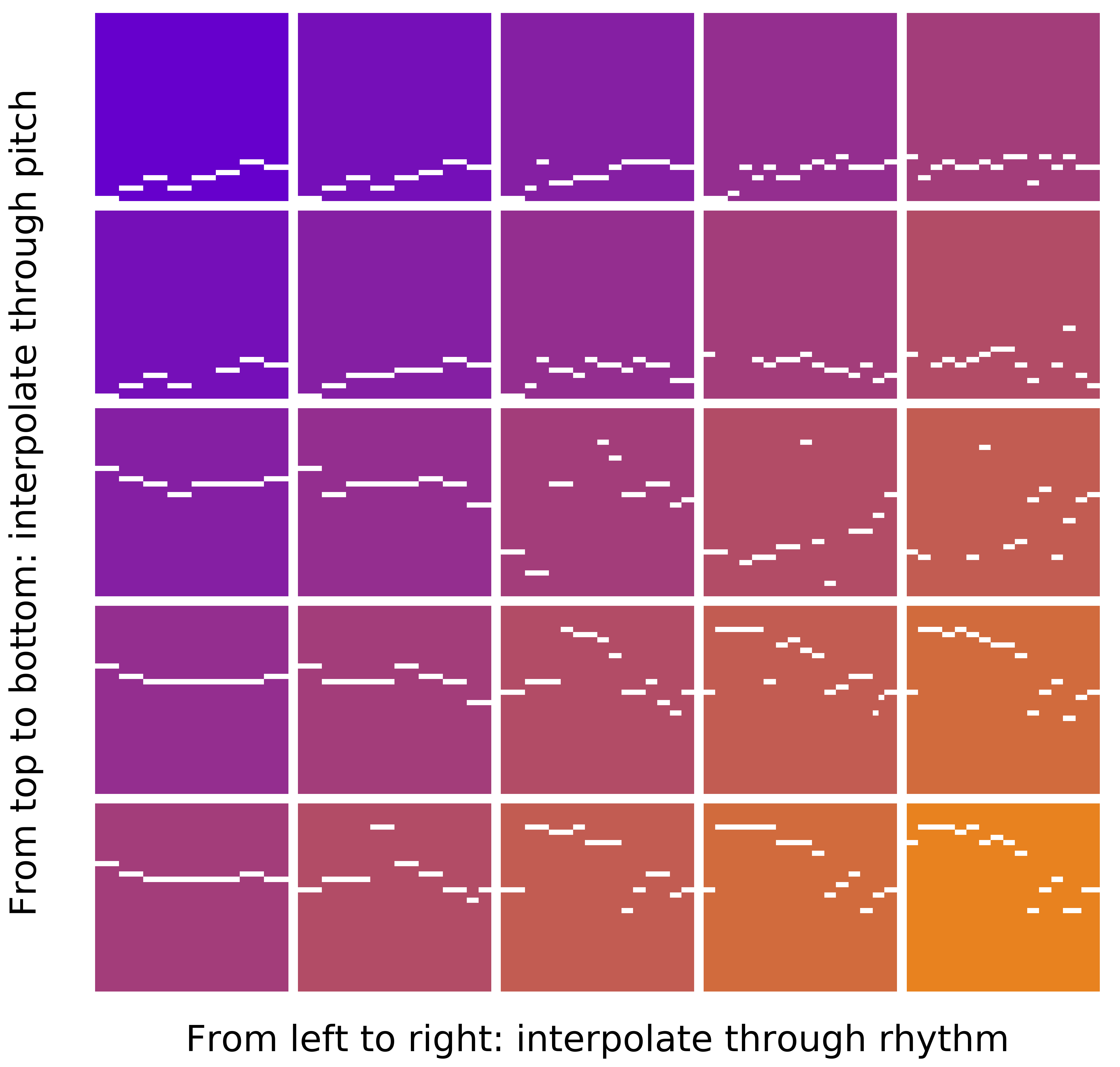}
	\caption{An illustration of two-way pitch-rhythm interpolation}
	\label{fig:interp}
\end{figure}

% \gus{let's make the writing easier}
Based on the pitch-rhythm disentanglement, the interface also enables a two-way interpolation from one music clip to another in a meaningful way associated with music style. Figure \ref{fig:interp} illustrates a grid view of the 2-way interpolation using the SLERP~\cite{watt1992advanced} method, where the top-left corner is the \textit{source} and the bottom-right one is the \textit{target}. The pitch interpolation is performed from the top to the bottom, the rhythm interpolation is performed from the left the the right, and the non-significant representations are also interpolated according to the Manhattan distances between a certain location to the target and source, respectively.

Note that the music mainly changes the rhythmic style horizontally and mainly changes the pitch style vertically. We can see this two-way interpolation as a powerful interface for composers to discover new music ideas with intuitive and precise controls on pitch and rhythm styles.
\\
\section{Conclusion and Future work}
\label{sec:conclusion-and-fugure-work}

In this paper, we inspect the latent space learned by a compressed MusicVAE model, and show that the space can be disentangled, with some dimensions corresponding to pitch change and some other dimensions corresponding to rhythm change. By tweaking values on those selected dimensions, we can modulate average pitch and rhythm complexity, or interpolate two music clips base on pitch or rhythm separately. We further design a user interface to control music creation by performing these operations. Our method still has some limitations. First, it cannot be used to determine the consequences of more complex operations on the latent representation. Second, our inspection is built through manual disentanglement. It would be helpful if an automatic disentangling model for music is designed and this should be a work in the future.

% The experiment result illustrates that different dimensions can take the charge of music concepts in latent representation. However, a important point that should not be ignored that dimensions are still correlated with each other. Our pitch analysis only shows how a music can have an overall rising or falling pitch; the rhythm is represented as a drum note. Thus, it's hard to demonstrate that the melody contour is fully controllable through manual disentanglement and inspection. Instead, it could be of great help by designing a automatic disentangling model for music, which means music creation could be more creative and reasonable by tweaking the parameters within the latent representation learnt by disentangling model.

% In this research, we demonstrate the melody midi sequences are controllable by learning their corresponding latent representation. Meanwhile, the overall pitch and rhythm could be controlled with only a few dimension within the latent representation. According to the experiment result, we design a intuitive demo interface for users to control music creation by directly tweaking the value on the latent representation or interpolating the corresponding dimensions between 2 totally different music clips.

% \section{Acknowledgement}
% Omitted for double-blind review.
%
% The following two commands are all you need in the
% initial runs of your .tex file to
% produce the bibliography for the citations in your paper.
\bibliographystyle{abbrv}

	\bibliography{nime-references.bbl}

\end{document}

%% file: math_commands.tex
\usepackage{amsmath,amsfonts,bm}

\def\eqref#1{equation~\ref{#1}}
\def\1{\bm{1}}

\def\ve{{\bm{e}}}

\def\vx{{\bm{x}}}

\def\vz{{\bm{z}}}

\DeclareMathAlphabet{\mathsfit}{\encodingdefault}{\sfdefault}{m}{sl}
\SetMathAlphabet{\mathsfit}{bold}{\encodingdefault}{\sfdefault}{bx}{n}
\newcommand{\tens}[1]{\bm{\mathsfit{#1}}}

\def\tM{{\tens{M}}}